\documentclass[letterpaper, 10 pt, conference]{ieeeconf}  % Comment this line out if you need a4paper
\usepackage{xcolor}
\usepackage{siunitx}
\usepackage{booktabs}
\usepackage{mathtools}
\usepackage{amssymb}
\usepackage{cite}
\usepackage{graphicx}
\usepackage{epstopdf}
\usepackage{subfig}
\usepackage{multirow}
\IEEEoverridecommandlockouts                              % This command is only needed if 
\title{\LARGE \bf
Lithium-ion Battery Capacity Prediction via Conditional Recurrent Generative Adversarial Network-based Time-Series Regeneration
}

\author{Myisha A. Chowdhury$^{1}$, Gift Modekwe$^{1}$, and Qiugang Lu$^{1,\dagger}$% <-this % stops a space
\thanks{*This work was supported by the Texas Tech University and NSF Grant 2340194.}% <-this % stops a space
\thanks{$^{1}$M.A. Chowdhury, G. Modekwe, and Q. Lu are with the Department of Chemical Engineering, Texas Tech University, Lubbock, TX 79405, USA.   
Email: {\tt\small myisha.chowdhury@ttu.edu; gmodekwe@ttu.edu; jay.lu@ttu.edu}}%
\thanks{$^{\dagger}$Corresponding author: Q. Lu.}%
}

\begin{document}

\maketitle
\thispagestyle{empty}
\pagestyle{empty}

%%%%%%%%%%%%%%%%%%%%%%%%%%%%%%%%%%%%%%%%%%%%%%%%%%%%%%%%%%%%%%%%%%%%%%%%%%%%%%%%
\begin{abstract}
Accurate capacity prediction is essential for the safe and reliable operation of batteries by anticipating potential failures beforehand. The performance of state-of-the-art capacity prediction methods is significantly hindered by the limited availability of training data, primarily attributed to the expensive experimentation and data sharing restrictions. To tackle this issue, this paper presents a recurrent conditional generative adversarial network (RCGAN) scheme to enrich the limited battery data by adding high-fidelity synthetic ones to improve the capacity prediction. The proposed RCGAN scheme consists of a generator network to generate synthetic samples that closely resemble the true data and a discriminator network to differentiate real and synthetic samples. Long short-term memory (LSTM)-based generator and discriminator are leveraged to learn the temporal and spatial distributions in the multivariate time-series battery data. Moreover, the generator is conditioned on the capacity value to account for changes in battery dynamics due to the degradation over usage cycles. The effectiveness of the RCGAN is evaluated across six batteries from two benchmark datasets (NASA and MIT). The raw data is then augmented with synthetic samples from the RCGAN to train LSTM and gate recurrent unit (GRU) models for capacity prediction. Simulation results show that the models trained with augmented datasets significantly outperform those trained with the original datasets in capacity prediction.
\end{abstract}

%%%%%%%%%%%%%%%%%%%%%%%%%%%%%%%%%%%%%%%%%%%%%%%%%%%%%%%%%%%%%%%%%%%%%%%%%%%%%%%%
\section{Introduction}
\label{sec:introduction}
Lithium-ion (Li-ion) batteries have gained widespread popularity for energy storage owing to their high energy density and long lifetime for numerous daily applications such as compact devices and electric vehicles  \cite{deng2015li}. Despite these advantages, the application of Li-ion batteries still faces challenges, particularly the capacity degradation over usage time, ultimately leading to the end-of-life of batteries, which is catastrophic for safety-critical systems such as electric vehicles \cite{zhang2023review}. Thus, how to accurately predict battery capacity and further state-of-health (SOH) proactively is crucial for the safety and reliability of battery-driven systems. 
%Li-ion batteries have challenges, such as performance degradation over time due to repetitive use and storage, ultimately causing battery capacity loss. However, an accurate estimate of the battery capacity is essential as it is a crucial metric to monitor the state of health (SOH) (current condition of the battery compared to the new one) and state of charge (SOC), i.e., the current charge level relative to its capacity,  of the battery.

Existing battery capacity prediction approaches can be categorized into model-based and data-driven methods \cite{chen2012prognostics}. Model-based methods focus on understanding the physical and chemical changes within the battery to give proper insight into how the degradation occurs. The main drawback of these models is the expensive computation due to the complicated electrochemical models involved. Additionally, they require detailed knowledge of the geometry and material properties of the battery, which may not always be available or easy to measure. On the other hand, data-driven methods use historical data collected during battery operation to predict the capacity \cite{zhang2023review}. They focus on identifying patterns and correlations between measured variables such as voltage, current, and temperature during charge/discharge cycles. By employing algorithms like neural networks (NNs), support vector machines, ensemble methods, and regression models, they can predict battery capacity and assess its SOH without the need for complex physical and chemical models \cite{li2019data, hannan2021deep}. Nonetheless, the effectiveness of these models relies on the quality and quantity of the available data \cite{li2019data}. However, owing to high experimental cost and data sharing restrictions, the amount of available data is limited in general \cite{qiu2023conditional}. Thus, data-driven methods often suffer from the \textit{``small data'' issue}.

Different data augmentation (DA) methods have been explored to overcome the above ``small data'' issue by artificially expanding the size and diversity of training datasets. Exemplary techniques include introducing random noise to bolster the robustness of data-driven models \cite{fan2022data} or using simulations to create data that replicates the variations observed in the real data \cite{channegowda2022attention}. Generative adversarial network (GAN) has emerged as a popular DA method, where an adversarial strategy is used for generating high-quality synthetic data \cite{goodfellow2014generative}. In contrast to other generative models, such as variational autoencoders, GANs are known for producing high-quality and realistic outputs that closely resemble the real samples with sharper and more detailed features \cite{makhzani2015adversarial}. %For this strategy, two players are involved in a competing way, where the generator tries to produce high-fidelity fake data whereas the discriminator tries to distinguish fake data from the real data. This adversarial way can force the generator to keep increasing the quality of fake data until they are indistringuisable 
%minimax game, where one contender tries to maximize an objective function while the other tries to minimize it []. 
However, for producing \textit{artificial time-series data} with GAN (as in the DA for batteries), special attention has to be paid to maintain the similarity in spatial and temporal correlations between the synthetic and real data. To this end, different variants of GANs, for instance, recurrent conditional GAN (RCGAN) \cite{esteban2017real} and time-series GAN \cite{yoon2019time}, have been developed by leveraging recurrent NN (RNN) and long short-term memory (LSTM) network for capturing temporal dynamics.

 In the context of DA for batteries, there is a growing interest in generating synthetic battery data using GAN. Battery data are inherently multivariate time-series, thereby demanding an understanding of the complex interplay between past and future samples, i.e., temporal correlations \cite{zhang2020time}. As a result, in \cite{naaz2021generative}, time-series GAN was implemented to synthesize the temperature, voltage, and state-of-charge (SOC) profiles. However, the generated voltage and SOC instances in \cite{naaz2021generative} show certain anomalies compared to the original data. Further, a hybrid method was proposed in \cite{wong2023novel}, where synthetic temperature, voltages, and current instances generated by GAN were used as inputs to an estimator for SOC calculation. In \cite{liu2024state}, a GAN-based reconstruction of missing voltage and aging features was proposed for capacity prediction. Despite the research progress witnessed, how to generate \textit{completely new charging/discharging cycle data} out of existing cycling data still remains a significant yet unsolved task for the DA of batteries. 

In this paper, we introduce a novel RCGAN scheme capable of generating high-quality new cycling data \textit{under unseen capacity values} while accurately capturing dynamic shifts resulting from aging and degradation of batteries. LSTM-based generator and discriminator are constructed to capture the temporal distributions alongside the spatial correlations of the raw data. One \textit{major novelty} of our method is that the generator is conditioned on the capacity value (decreases as battery cycles continue) to learn the changes in battery dynamics over charging-discharging cycles. The proposed model is tested on the benchmark NASA and MIT battery datasets \cite{saha2007battery,severson2019data}. Extensive assessment is conducted to validate the effectiveness of the trained RCGAN and the quality of the synthetic samples. Completely new cycling data under unseen capacity values are then generated to augment the raw data, and it is shown that the battery capacity prediction can be significantly improved with DA. 

 %Note that by conditioning the generator alone, we address an inherent learning issue of typical GANs caused by the discriminator overpowering the generator due to the disparity in the difficulty of the corresponding tasks \cite{goodfellow2016nips}.

\section{Preliminaries}
\label{sec:preliminaries} 

\subsection{Generative adversarial network (GAN)}
\label{subsec:gan}

GANs are a subclass of generative models, first proposed by Goodfellow et al. \cite{goodfellow2016nips,goodfellow2014generative}, to generate data instances without relying on a pre-defined hypothesis. Specifically, standard GAN consists of two networks: a generator and a discriminator \cite{gui2021review}. The generator $G(\cdot)$, parameterized by $\theta_g$, produces synthetic samples $x$ by mapping random noise $z\sim p_{z}(z)$ to the data space: $x=G(z)$. Meanwhile, the discriminator $D(\cdot)$, parameterized by ${\theta_d}$, provides a scalar output, $D(x) \in [0,1]$, describing the probability of $x$ belonging to the real data. The discriminator is trained to maximize its ability to accurately distinguish between real data sampled from the training distribution $p_{data}(\cdot)$ and the synthetic examples produced by the generator. In contrast, the generator attempts to generate synthetic data highly similar to real data instances to deceive the discriminator. In other words, the generator and discriminator are engaged in a minimax game, where the former tries to minimize while the latter attempts to maximize the following function:
\begin{align}
\min_G  \max_D  V(D, G)  & = \mathbb{E}_{x \sim p_{data}(x)}\left[\log D(x)\right]+ \nonumber \\ & \mathbb{E}_{z \sim p_z(z)}\left[\log(1-D(G(z)))\right],
\label{eq:standard_gan}
\end{align}
where $\log D(x)$ is the cross entropy between $[1 \quad 0]^T$ and $[D(x) \quad 1-D(x)]^T$, and $\log(1-D(G(z)))$ is resulted from the cross entropy between $[0 \quad 1]^T$ and $[D(G(z)) \quad 1-D(G(z))]^T$ (see \cite{gui2021review} for details).

\subsection{Conditional GAN (CGAN)}
\label{subsec:cgan}
Conditional GAN is one of the most popular variants of GAN, where both discriminator and generator receive additional labels $c$, known as conditioning features \cite{ding2020ccgan}. This conditioning allows the control over the modes at which the data is generated. Unlike standard GANs, CGAN not only attempts to generate samples similar to real data instances but also tries to match the corresponding labels. The objective function of the CGAN can be formulated as \cite{ding2020ccgan}
\begin{align}
\min_G  \max_D V(D, G) & = \mathbb{
E}_{x \sim p_{data}(x)}\left[\log D(x|c)\right]+ \nonumber \\ &\mathbb{E}_{z \sim p_z(z)}\left[\log(1-D(G(z|c)))\right].
\label{eq:cgan}
\end{align}

\section{Methodology}
\label{section:methodology}
\subsection{RCGAN for synthetic battery data}
\label{subsection:rcgan}
This section introduces the RCGAN model to generate synthetic samples that capture the spatial correlations and temporal dynamics of the multivariate time-series battery profiles. The model is trained using limited real battery data instances.

\subsubsection{Data preparation}
\label{subsection:preprocessing}
Before passing the data into the RCGAN, we preprocess it to ensure uniformity. First, we downsample the battery profile sequence in each cycle $k\in\mathcal{K}:=\{1,\ldots,K\}$ to length $l$ for faster training, where $K$ is the total number of available cycles in training data. Moreover, min-max standardization is applied to the input features, i.e., voltage $V_t^{[k]}$, temperature $T_t^{[k]}$, and current $I_t^{[k]}$, $\forall t\in \mathcal{I}:=\{1,\ldots,l\}$, of each cycle $k$ to ensure that their amplitudes are within $(-1,1)$: 
\begin{equation}
\tilde{x}^{[k]}_{t} = \frac{2(x^{[k]}_t-\underline{x}^{[k]})}{\bar{x}^{[k]}-\underline{x}^{[k]}} - 1,~\forall t\in \mathcal{I},~~\forall k\in\mathcal{K}, 
\label{eq:minmax}
\end{equation}
where $x_{t}^{[k]}\in\{V_t^{[k]},T_t^{[k]},I_t^{[k]}\}$, and  $\bar{x}^{[k]}$ and $\underline{x}^{[k]}$ are the maximum and minimum values of $x_{t}^{[k]}$ in cycle $k$. 

\subsubsection{Architecture of RCGAN}
\label{subsection:architecture}
One crucial aspect of GAN for synthesizing time-series data is that along with capturing feature distributions at individual time points, it shall also encapsulate the potentially intricate dynamics of those variables across different time steps. To achieve this, we incorporate two LSTM layers into the generator $G(\cdot)$, followed by a fully connected output layer of dimension $3\times l$, i.e., $G(\cdot)$ produces the \textit{entire synthetic cycle} profiles at once: $\forall k\in\mathcal{K}$,
\begin{equation}
\hat{x}^{[k]}:=[\hat{V}^{[k]},\hat{I}^{[k]},\hat{T}^{[k]}]=G(\hat{z}^{[k]}|c^{[k]})\in\mathbb{R}^{3\times l},\label{eq:GAN_archi}
\end{equation}
where $\hat{z}^{[k]}=[z^{[k]}_1,\ldots,z^{[k]}_l]$,  $\hat{V}^{[k]}=[V^{[k]}_1,\ldots,V^{[k]}_l]\in\mathbb{R}^{l}$ and the same holds for $\hat{I}^{[k]}$ and $\hat{T}^{[k]}$. Random noise $z^{[k]}_t \sim p_{z}(z), \forall t\in\mathcal{I}$. The LSTM layers capture the dynamics across time steps, allowing the generator to create time-series preserving the original dynamics in the raw data. All layers use $tanh(\cdot)$ as the activation function. The synthetic profiles $\hat{x}^{[k]}$ (labeled 0) at cycle $k$ and the real profiles $x^{[k]}\in\mathbb{R}^{3\times l}$ (labeled 1) are jointly sent to the  discriminator $D(\cdot)$ as inputs. The role of the discriminator is to distinguish between real and synthetic samples. The architecture of the discriminator closely resembles that of the generator, except for the output layer, which has a single unit (instead of a vector as for the generator). We use $tanh(\cdot)$ and $sigmoid(\cdot)$ as the activation functions for LSTM and output layers, respectively.

Note that battery data is \textit{cyclic} in nature, i.e., the battery time-series profiles are generated according to similar protocols for each cycle. However, they also differ between cycles due to battery aging and degradation. The capacity $c^{[k]}$ (at cycle $k$) is a critical SOH indicator for explaining the shift in battery behaviors due to aging. Therefore, using capacity as a conditioning factor in \eqref{eq:GAN_archi} can assist the generator in capturing changes in battery dynamics over cycles. Nonetheless, during operation, batteries often have \textit{capacity regeneration} stages, where a sudden increase in available capacity is observed after a complete charging-discharging cycle \cite{pan2022method}. These spikes can mislead the generator because other variables like temperature, voltage, and current behave similarly in those regions despite the high capacity value. To address this issue, we smooth out the capacity values using moving mean
\begin{equation}
\hat{c}^{[k]}=\frac{1}{2m+1}\sum_{i=k-m}^{k+m}c^{[i]},~k\in\{m+1,\ldots,K-m\},
\end{equation}
where $\hat{c}^{[k]}$ for cycles outside the above set remains the same as the raw capacity values, and $2m+1$ is the window size. The smoothed capacity $\hat{c}^{[k]}$ decreases monotonically and will be used as the conditioning feature for the generator. 

\subsubsection{Training of RCGAN}
\label{subsec:training}

The balance between the generator and discriminator during GAN training is crucial to ensure convergence \cite{goodfellow2016nips}. However, maintaining this balance is particularly challenging because of the disparate tasks assigned to each component. The generator faces a more daunting task compared to the discriminator. The generator synthesizes lengthy, multivariate time-series data of a given length $l$, whereas the discriminator's role is relatively simple: distinguishing between real and synthetic data samples. Consequently, the discriminator often outperforms the generator, leading to issues such as the generator's gradients becoming too small, thereby impeding the training process \cite{goodfellow2016nips}.

To mitigate this problem, we depart from the conventional CGAN approach: instead of conditioning both generator and discriminator on external factors, we exclusively\textit{ condition only the generator} on the capacity. This decision aims to give the generator an advantage over the discriminator, assisting to the balance of the training process on both sides. Introducing the smoothed capacity $\hat{c}$ of a \textit{generic cycle}  as conditioning into the generator $G(z|\hat{c})$, the new objective for training the networks in RCGAN is formulated as
\begin{align}
\min_G \max_D V(D, G) & = \mathbb{E}_{x \sim p_{data}(x)}\left[\log D(x)\right]+ \nonumber\\
&\mathbb{E}_{z \sim p_z(z)}\left[\log(1-D(G(z|\hat{c})))\right].
\label{eq:cgan_new}
\end{align}
Consider $K$ cycles of available training data each consisting of $n$ time-steps. The expectations in \eqref{eq:cgan_new} are replaced by empirical means:
\begin{align}
\min_{\theta_{g}} \max_{\theta_{d}} V(\theta_{g}, \theta_{d}) & = \frac{1}{K}\sum\nolimits_{k=1}^{K}\sum\nolimits_{t=1}^{n}\left[\log D_{\theta_d}(x^{[k]}_t)+ \nonumber \right.\\ 
&\left.  \log[1-D_{\theta_d}(G_{\theta_g}(\hat{z}^{[k]}_t|\hat{c}^{[k]}))] \right].
\label{eq:cgan_new_empirical}
\end{align}
We will employ the ADAM optimizer to solve the optimization problem in \eqref{eq:cgan_new_empirical}, and define optimized parameters as $\theta_{g}^{*}$ for the generator and $\theta_{d}^{*}$ for the discriminator.

%Moreover, the discriminator and the generator are trained alternatively to create a stable training process where each network can focus on its specific task without interference. In other words, the generator network is kept frozen during the discriminator training. This makes the discriminator's task clearer as it must distinguish between real data and data generated by a fixed, unchanging generator. Conversely, during generator training, the discriminator is kept constant. If the discriminator were to change during generator training, it would face a shifting target. This could lead to difficulties in convergence, as the generator would need to adapt its output to match the changing discriminator continuously.

\subsection{DA for capacity prediction of batteries}
\label{subsec:soh_estimation}

\begin{figure*}[tbh]
\centering
\includegraphics[width=2\columnwidth]{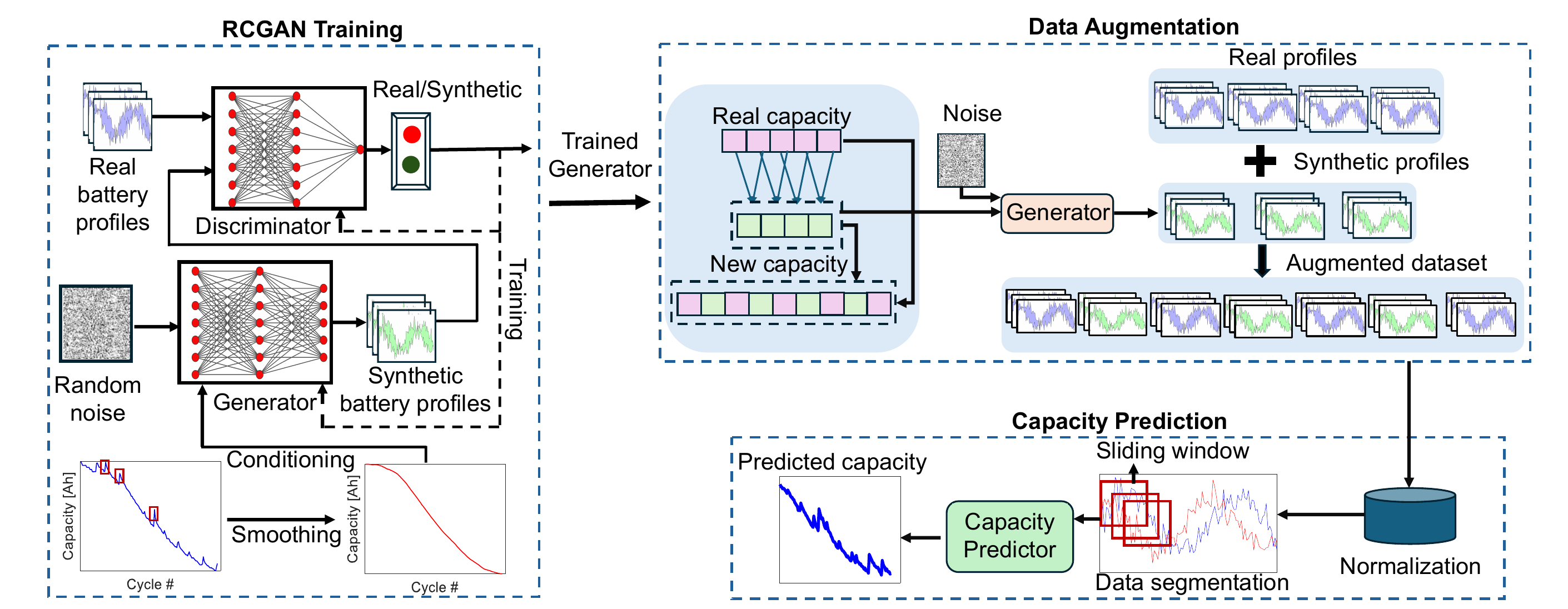}
\caption{Schematic illustrating the process of training and applying RCGAN for augmenting battery datasets. The trained RCGAN model generates new cycle data to enhance the dataset, ultimately improving battery capacity prediction.} \label{fig:idea_flow}
\end{figure*}
After training the RCGAN model, we generate synthetic profiles by creating \textit{new cycles under unseen capacities} and augment the original dataset to enhance its diversity. For a given new capacity value (after smoothing) $\hat{c}^{\prime}$, the synthetic cycle profiles can be obtained as
\begin{equation}
[\hat{V}^{\prime},\hat{I}^{\prime},\hat{T}^{\prime}]=G_{\theta_g^{*}}(\hat{z}^\prime|\hat{c}^{\prime}),~z_t^{\prime}\sim p_{z}(z),~t\in\mathcal{I}. 
\end{equation}
Theoretically $\hat{c}^{\prime}$ can be freely chosen. However, one practical option is the averaged capacity of neighboring two cycles of the training data, such that a new cycle is created and added between every two cycles. Then the raw and synthetic training data are integrated for DA and capacity prediction. Fig. \ref{fig:idea_flow} shows the entire DA and capacity prediction process. 

In this work, we train LSTM and gated recurrent unit (GRU) models for capacity prediction. For comparison, both models are trained on the raw training and augmented training data. After training, future capacities in the test data are predicted and assessed. We use root mean square error (RMSE) and mean absolute error (MAE) as the metrics for assessing the prediction performance. The pseudo code of the entire algorithm is given in Table  \ref{table:pseudocode}.

\begin{table*}[tbh]	
\centering
\caption{Pseudo code for battery capacity prediction with RCGAN-based DA.} \label{table:pseudocode}
\scalebox{1}{
\begin{tabular}{ll}
\hline
\multicolumn{2}{l}{\textbf{Algorithm: Capacity prediction with RCGAN-based DA}} \\ \hline
1: & \textbf{Input}: Discriminator: Real and synthetic time-series cycle data $x^{[k]}$ and $\hat{x}^{[k]}$, $\forall k\in\mathcal{K}:=\{1,\ldots,K\}$; \\
& \quad \quad \quad Generator: Noise vector $\hat{z}^{[k]}$ and smoothed capacity $\hat{c}^{[k]}$ as conditioning factor, $\forall k\in\mathcal{K}$;\\
2: & Initialize the discriminator and generator parameters $\theta_d$ and $\theta_g$, respectively;\\
3. & \textbf{for} $i=1,\ldots,epoch ~duration$ \textbf{do} \\
4: & \quad  Sample a random batch of $s$ smoothed capacities $\hat{c}^{[s]} \in \mathbb{R}^s$ from $\{\hat{c}^{[1]},\ldots,\hat{c}^{[K]}\}$,  and sample noise vectors $\hat{z}^{[s]} \in \mathbb{R}^{s \times l \times d}$ \\&  \quad    from prior distribution $p_z(\cdot)$, where $l$ is the length of each cycle, and $d$ is the input size for LSTM in $G_{\theta_g}(\cdot)$;\\
5: & \quad if $i \% 3=0$, update the discriminator network: \\
6. & \quad \quad \quad Generate synthetic samples $\hat{x}^{[s]} \in \mathbb{R}^{s \times l \times 3} $ with generator $G_{\theta_g}(\hat{z}|\hat{c})$ and sample $s$ cycles of real data $x^{[s]}$ corresponding to $\hat{c}^{[s]}$;\\
7. & \quad \quad \quad Update the discriminator parameter $\theta_d$ using stochastic gradient \textbf{ascent}: \\
& \quad \quad \quad  \qquad $\theta_d \leftarrow \theta_d +\nabla \bar{V}_d(\theta_d)$, where $\bar{V}_d(\theta_d) = \frac{1}{s}\sum_{i=1}^{s} \left[ \log D_{\theta_d}(x^{[i]})+\log(1-D_{\theta_d}(\hat{x}^{[i]}))\right]$;\\
8. & \quad else, update the generator network:\\
9. & \quad \quad \quad Update the generator parameter $\theta_g$ using stochastic gradient \textbf{descent}:  \\
& \quad \quad  \quad \qquad $\theta_g \leftarrow \theta_g -\nabla \bar{V}_g(\theta_g)$, where $\bar{V}_g(\theta_g) = \frac{1}{s}\sum_{i=1}^{s} \left[\log(1-D_{\theta_d}(G_{\theta_g}(\hat{z}^{[i]}|\hat{c}^{[i]})))\right]$;\\
10. & end for \\
11. & Repeat the above loop until $\theta_g$ and $\theta_d$ converge to $\theta_g^*$ and $\theta_d^*$, respectively;\\
12. & \textbf{After training is completed:}\\
13. & Compute unseen capacity vector $\hat{c}^{[k]^\prime}$ by averaging out the capacities of neighboring two cycles in the training data:\\
& \qquad $\hat{c}^{[k]^\prime}=\left(\hat{c}^{[k]}+\hat{c}^{[k+1]}\right)/2$, $\forall k\in \mathcal{K}^\prime:=\{1,\ldots,K-1\}$; \\
14. & Sample $\hat{z}^{[k]^\prime} \in \mathbb{R}^{ l\times d}\sim p_{z}(\cdot), \forall k\in \mathcal{K}^\prime$, generate new cycle data with the trained generator: $\hat{x}^{[k]^\prime}=G_{\theta_g^*}(\hat{z}^{[k]^\prime}| \hat{c}^{[k]^\prime}), \forall k\in \mathcal{K}^\prime$;\\ 
15. & Augment the new cycle data with the existing data: $X=\{x^{[1]},\ldots,x^{[K]}\}\cup\{\hat{x}^{[1]^{\prime}},\ldots,x^{[K-1]^\prime}\}$;\\
16. & Leverage the augmented dataset $X$ to train LSTM- and GRU-based capacity prediction models;\\
17. & Use the trained LSTM and GRU models to predict the capacity values over the $n$ test cycles. \\
\hline
\end{tabular}}%
\end{table*}

\section{Result and discussion}
\label{sec:result}
We use the benchmark NASA battery (NB) \cite{saha2007battery} and MIT battery (MB) datasets \cite{severson2019data} to validate the proposed RCGAN-based DA and capacity prediction methods. Three batteries from each set, i.e., 5, 6, and 7 from NB and 1, 5, and 7 from the MB, are selected in this study. For the three NASA batteries, we use all the \textit{charging cycles}, and for the three MIT batteries, we use the first 150 \textit{charging cycles} of each battery for the experiments below. 

\subsection{Training and evaluation of the RCGAN model}
\subsubsection{Training curve}
\begin{figure}[tbh]
\centering
\includegraphics[width=\columnwidth]{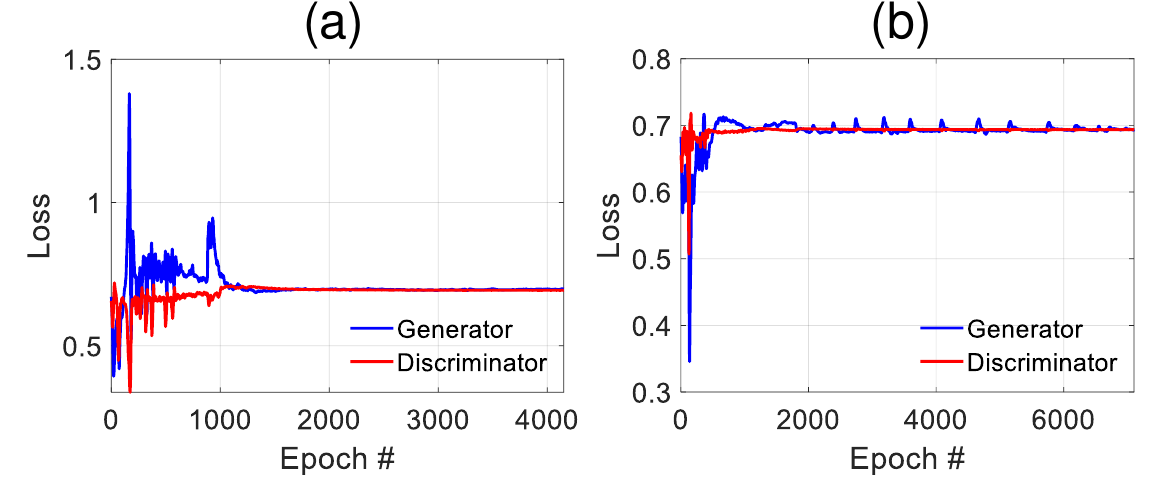}\\
\caption{Loss curves of the generator and discriminator for (a) NB \#5 and (b) MB \#1 during training.
} \label{fig:loss}
\end{figure}
The first $K=100$ cycles of each battery data are used to train the generator and discriminator in the RCGAN. 
%To train the generator and discriminator of the RCGAN, we use the first 100 cycles from the battery data. We expect the quality of the synthetic data to improve as the training progresses (decrease in the loss). 
Fig. \ref{fig:loss} demonstrates the loss curves of the generator and discriminator with NB \#5 and MB \#1 throughout the training process. The curves show that the generator and the discriminator compete against each other as they attempt to minmax the loss function in \eqref{eq:cgan_new}. In other words, as the performance of one network improves, the performance of the other network decreases. At the beginning, the loss of the generator is higher than that of the discriminator for a significant number of epochs. This difference indicates that the discriminator is proficient at distinguishing between real and synthetic samples at the beginning, i.e., the generator faces difficulty in fooling the discriminator. As the training proceeds, the generator improves its ability to generate samples that closely resemble the real data, and the discriminator faces a more challenging task of distinguishing between real and generated samples. Towards the end of training, both networks stabilize and converge to equilibrium. A similar learning behavior is observed for the other 4 batteries under study. This convergence illustrates the ability of the RCGAN model to effectively learn and capture the underlying dynamics in the raw data, thereby demonstrating its robustness and adaptability across different battery datasets and operating conditions.  

%For training the models using Eq. \eqref{eq:cgan_new}, adaptive moment estimation optimizer, commonly known as ADAM, is leveraged with learning rate $1.4 \times 10^{-5}$ and  $5 \times 10^{-5}$ for the discriminator and generator, respectively.

\subsubsection{Evaluation of the synthetic data}
\begin{figure*}[tbh]
\centering
\includegraphics[width=1.75\columnwidth]{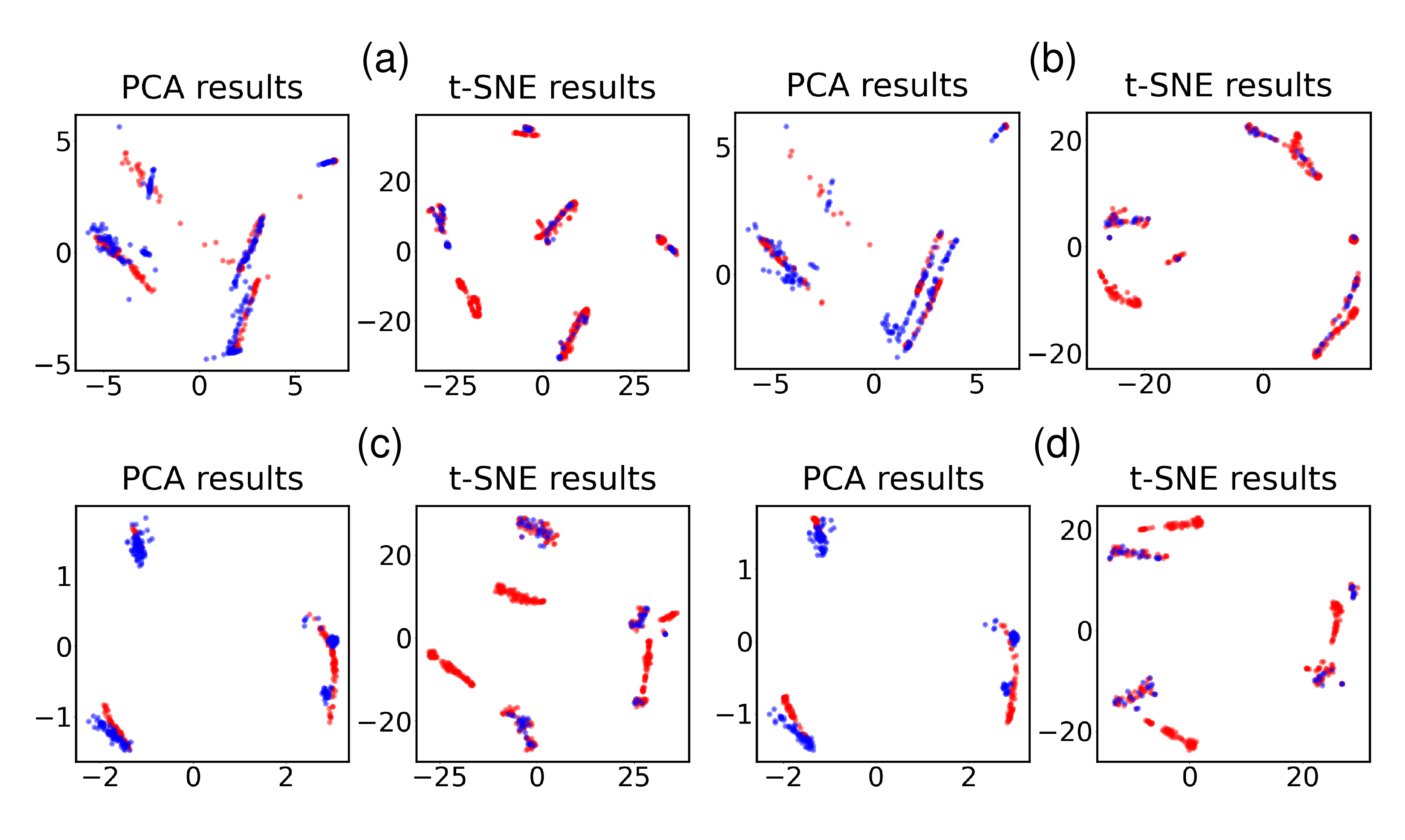}
\caption{Dimension reduction with PCA and t-SNE for real (blue) and synthetic (red) samples of NB \#5 (top) and MB \#1 (bottom) for training cycles (a, c) and test cycles (b, d).} \label{fig:pca_tsne}
\end{figure*}
As stated before, we split each battery data into the first $K$ training cycles and the remaining $n$ test cycles, where $(n+K)$ is the total cycle number. Battery aging information is provided to the generator during training by conditioning it on smoothed capacities $\hat{c}^{[1:K]}$. We then use the trained generator to generate synthetic profiles for both observed training cycles and unseen test cycles. This evaluation procedure allows us to understand the generalization capability of the trained model, as it generates synthetic cycles for unseen future capacity conditions beyond those encountered during training. We perform dimensionality reduction using principal component analysis (PCA) and t-distributed stochastic neighbor embedding (t-SNE) to visualize the similarity between the high-dimensional real and synthetic data for the training and test cases. Fig. \ref{fig:pca_tsne} shows a comprehensive visualization, revealing a substantial overlap between the synthetic (red) and real (blue) instances for NB \#5 (top) and MB \#1 (bottom) for both training (a, c) and test (b, d) sets. This overlap suggests that the synthetic data closely replicate the intricate characteristics and distributions present in the real samples. We then randomly select temperature, voltage, and current profiles from synthetic datasets and compare them with original data under the same capacity.
\begin{figure*}[tbh]
\centering
\includegraphics[width=1.75\columnwidth]{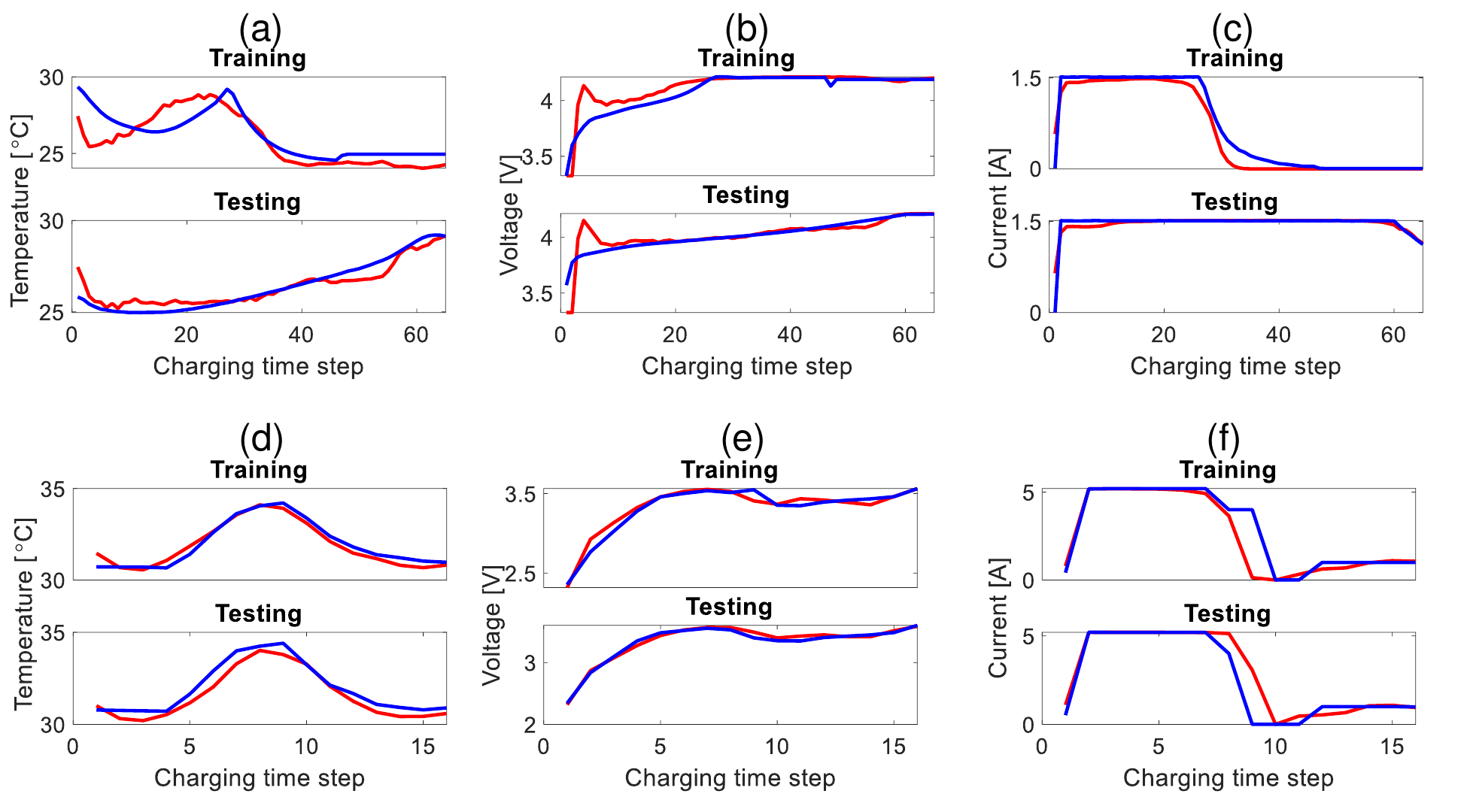}
\caption{Comparison between real (blue) and synthetic (red) cycle profiles of temperature, voltage, and current for NB \#5 (a-c) and MB \#1 (d-f).} \label{fig:extrapolation_figure}
\end{figure*}
The results showcased in Fig. \ref{fig:extrapolation_figure} underline the performance of the generator. Notably, the generator not only synthesizes temperature, voltage, and current profiles for capacity values already encountered during training (denoted as ``training"), but also generates high-fidelity profiles for unseen capacities in the last $n$ test cycles (designated as ``testing"). This observation underscores the robustness and versatility of the generator in extrapolating beyond the observed data, effectively capturing the aging-dependent dynamic patterns in the original data. 

\begin{figure*}[tbh]
\centering
\includegraphics[width=1.8\columnwidth]{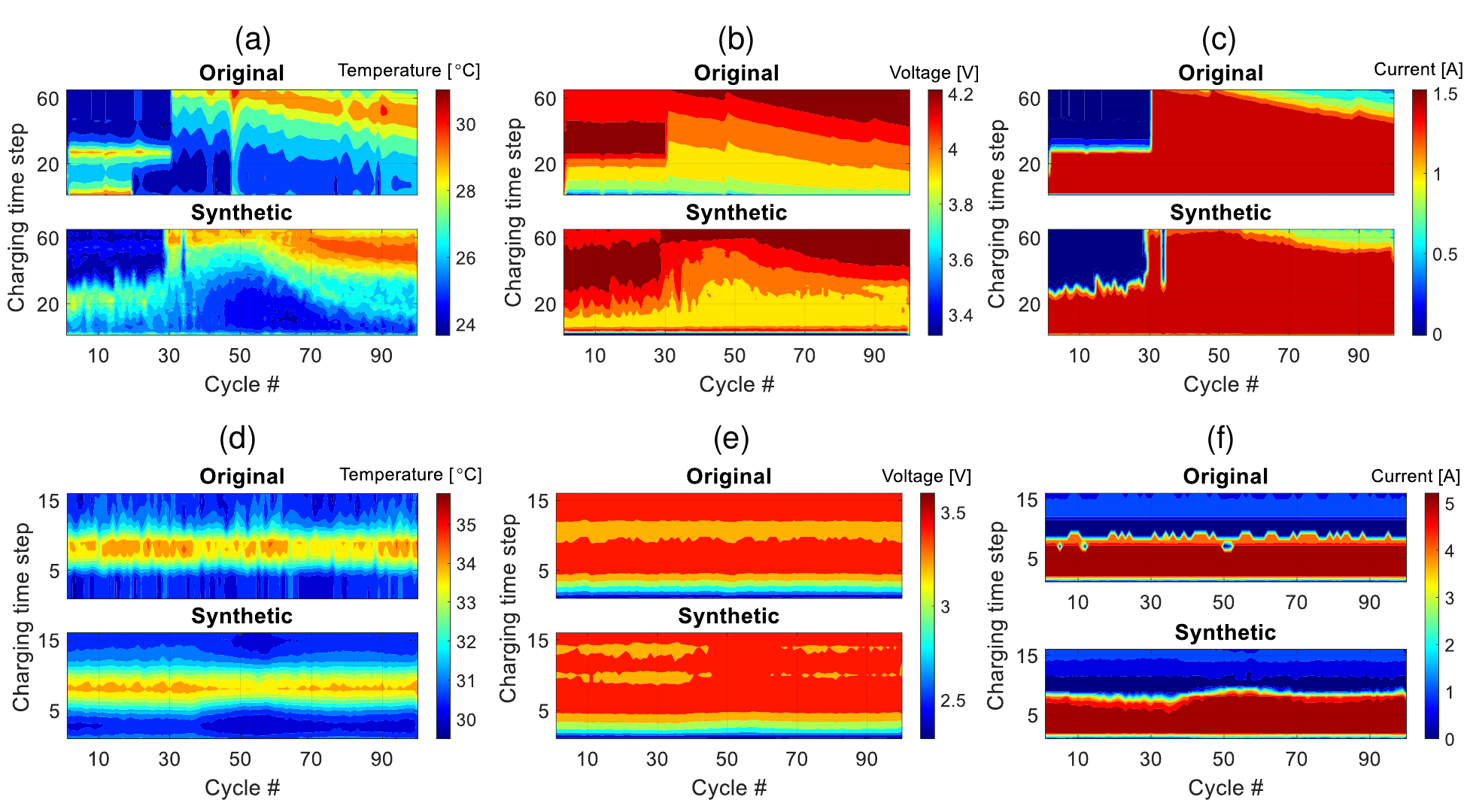}
\caption{Comparison between battery profiles (temperature, voltage, and current) of different cycles from the raw training and RCGAN-based synthetic (under new capacity values $\hat{c}^{[k]^\prime}$) data for NB \#5 (a-c) and MB \#1 (d-f).} \label{fig:contour_plot}
\end{figure*}
After evaluating the generator's performance, we augment the existing battery datasets by creating new cycle data under capacity values \textit{not seen in both training and test cycles}. As in Section \ref{subsec:soh_estimation}, practically, we introduce a new cycle between every two existing cycles in the raw training data, where the unseen capacity $\hat{c}^{[k]^\prime}=\left(\hat{c}^{[k]}+\hat{c}^{[k+1]}\right)/2$, $\forall k\in\{1,\ldots,K-1\}$. These new capacities serve as conditioning features for the generator, enabling the generation of synthetic cycle data. Fig. \ref{fig:contour_plot} compares the battery profiles from raw training cycles with the synthetic profiles from new cycles with capacity $\hat{c}^{[k]^\prime}$, for NB \#5 (a-c) and MB \#1 (d-f). Similar results are observed in the other 4 batteries.

\subsection{Capacity prediction with augmented battery data}

\begin{figure}[!t]
\centering
\includegraphics[width=\columnwidth]{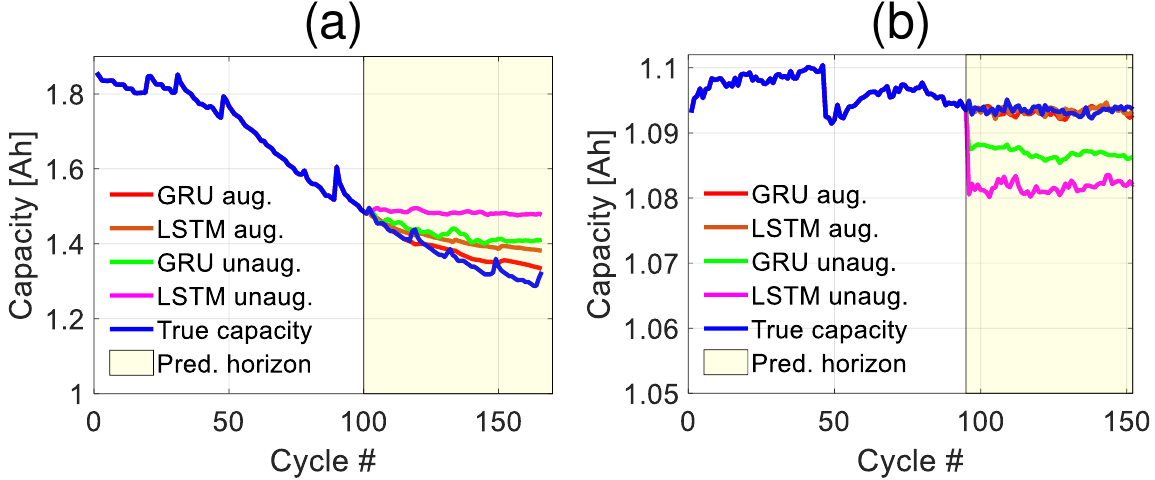}
\caption{Capacity predictions from GRU and LSTM trained on augmented and original (a) NB \#5 and (b) MB \#1 data. 
}\label{fig:capacity_prediction}
\end{figure}

After the RCGAN-based DA, the augmented dataset consisting of $K$ original and $K$ synthetic cycling data (temperature, voltage, and current profiles over cycles) are used to train the GRU and LSTM models for capacity prediction. Fig. \ref{fig:capacity_prediction} shows the performance of the LSTM and GRU-based capacity forecast on NB \#5 and MB \#1 batteries with both original and augmented datasets. The results indicate that network models trained on augmented datasets clearly outperform those on the original ones in the capacity prediction. This suggests that the synthetic data instances closely align with the underlying real data distribution, and can enrich the raw data and further enhance the capacity prediction performance. Additionally, the incorporation of the new cycle data mitigates the risk of overfitting, thereby further bolstering the performance. Similar results are observed in other batteries as provided in Table \ref{table:error}.
\begin{table}[tbh]
\centering
\caption{Capacity prediction errors of different batteries.}
\label{table:error}
\begin{tabular}{llllll}
\hline
Battery & \multirow{2}{*}{Criteria} & GRU  & GRU  & LSTM  & LSTM  \\
Name &  & Original & Augmented & Original & Augmented \\
\hline
\multirow{2}*{NB \#5} & RMSE & 0.0681 & \textbf{0.0252} & 0.1250 & 0.0534\\
& MAE & 0.0589 & \textbf{0.0198} & 0.1146 & 0.0452\\
\hline
\multirow{2}*{NB \#6}  & RMSE & 0.1258 & \textbf{0.0341} & 0.1009 & 0.0640 \\
& MAE & 0.1138 & \textbf{0.0291}  & 0.0876 & 0.0517 \\
\hline
\multirow{2}*{NB \#7} & RMSE & 0.0804  & \textbf{0.0201} & 0.1492 & 0.0537 \\
& MAE & 0.0698 & \textbf{0.0056} & 0.1419 & 0.0328 \\
\hline
\multirow{2}*{MB \#1}  & RMSE & 0.0067 & \textbf{0.0008} & 0.0148 &  \textbf{0.0008} \\
& MAE & 0.0066 &  \textbf{0.0007} & 0.0148 & \textbf{0.0007} \\
\hline 
\multirow{2}*{MB \#5} & RMSE & 0.0034 & \textbf{0.0007} & 0.0089 & 0.0012 \\
& MAE & 0.0033 & \textbf{0.0006}  & 0.0089 & 0.0009 \\
\hline
\multirow{2}*{MB \#7} & RMSE & 0.0257 & \textbf{0.0011} & 0.0285 & 0.0065 \\
& MAE & 0.0257 & \textbf{0.0009} & 0.0285 &  0.0062\\
\hline
\end{tabular}%
\end{table}

Overall, the GRU model exhibits superior capacity prediction performance compared to the LSTM model for both augmented and original datasets. This disparity may be attributed to LSTM's susceptibility to overfit due to its intricate architecture \cite{yamak2019comparison}. GRUs, with their simplified gating mechanism, may generalize better to the battery dataset, allowing for more effective learning of long-term dependencies while mitigating the risk of overfitting. Moreover, it is observed that the error in the capacity prediction increases across all methods as the cycle number progresses. This inclination in error can be attributed to the growing uncertainty associated with the battery's condition over successive cycles. Despite this increase in error, the relative performance of the trained GRU based on RCGAN-augmented data remains satisfactory across the entire prediction horizon.

\section{Conclusions}
	This work presents a GAN framework named RCGAN to synthesize multivariate time-series battery data for enhanced capacity prediction. The proposed RCGAN employs an LSTM-based generator and discriminator to capture both temporal and spatial distributions in the data. Further, we condition the generator on the capacity value to enable its incorporation of battery aging characteristics. This framework can enhance the fidelity of synthetic data by capturing the complexities in the battery cycling profiles that are also dependent on the capacity (aging). The proposed RCGAN model is assessed with the benchmark NASA and MIT datasets. Simulation results show that the trained RCGAN model can produce synthetic data closely resembling the real instances across all test batteries, for cycling profiles under both observed and unseen capacity values. Moreover, LSTM and GRU models are trained, based on both raw and augmented datasets with RCGAN, to predict future capacities. It shows that with DA, the capacity prediction can be significantly improved compared with the models trained without DA in all cases. In the future work, we will evaluate the generalizability of the RCGAN model by training and testing it on separate battery datasets. Additionally, we will utilize the generated samples for SOC and state-of-power (SoP) estimation to enhance battery prognostics.
\section{Acknowledgment}
M. A. Chowdhury acknowledges the support of Distinguished
Graduate Student Assistantships (DGSA) from Texas Tech University. Q. Lu acknowledges the new faculty startup funds from Texas Tech University. The authors acknowledge the support from the National Science Foundation under Grant No. 2340194.

\bibliographystyle{ieeetr}
\bibliography{gan_battery}

\end{document}